\documentclass[journal]{IEEEtranTIE}
\usepackage{graphicx}
\usepackage{cite}
\usepackage{picinpar}
\usepackage{amsmath}
\usepackage{url}
\usepackage{flushend}
\usepackage{colortbl}
\usepackage{soul}
\usepackage{multirow}
\usepackage{pifont}
\usepackage{color}
\usepackage{alltt}
\usepackage[hidelinks]{hyperref}
\usepackage{enumerate}
\usepackage{siunitx}
\usepackage{epstopdf}
\usepackage{pbox}
\usepackage{amsmath,amssymb,amsfonts}

\usepackage{algorithmic}
\usepackage{graphicx}
\usepackage{textcomp}
\usepackage{xcolor}
\usepackage{cuted}
\usepackage{stfloats}
\usepackage[RPvoltages]{circuitikz}
\usepackage[thinc]{esdiff}
\usepackage{tikz}
\usetikzlibrary{arrows,matrix,positioning,fit}
\usetikzlibrary{calc}
\usetikzlibrary{shapes}
\ctikzset{bipoles/cuteinductor/coils=10}
\ctikzset{bipoles/cuteinductor/width=1.5}
\def\BibTeX{{\rm B\kern-.05em{\sc i\kern-.025em b}\kern-.08em
    T\kern-.1667em\lower.7ex\hbox{E}\kern-.125emX}}
\usepackage[export]{adjustbox}
\usepackage{mathrsfs}
\def\BibTeX{{\rm B\kern-.05em{\sc i\kern-.025em b}\kern-.08em
    T\kern-.1667em\lower.7ex\hbox{E}\kern-.125emX}}
\usepackage{scalerel}
\usepackage{siunitx}
\usetikzlibrary{svg.path}
\definecolor{orcidlogocol}{HTML}{A6CE39}
\tikzset{
  orcidlogo/.pic={
    \fill[orcidlogocol] svg{M256,128c0,70.7-57.3,128-128,128C57.3,256,0,198.7,0,128C0,57.3,57.3,0,128,0C198.7,0,256,57.3,256,128z};
    \fill[white] svg{M86.3,186.2H70.9V79.1h15.4v48.4V186.2z}
                 svg{M108.9,79.1h41.6c39.6,0,57,28.3,57,53.6c0,27.5-21.5,53.6-56.8,53.6h-41.8V79.1z M124.3,172.4h24.5c34.9,0,42.9-26.5,42.9-39.7c0-21.5-13.7-39.7-43.7-39.7h-23.7V172.4z}
                 svg{M88.7,56.8c0,5.5-4.5,10.1-10.1,10.1c-5.6,0-10.1-4.6-10.1-10.1c0-5.6,4.5-10.1,10.1-10.1C84.2,46.7,88.7,51.3,88.7,56.8z};
  }
}
\newcommand\orcidicon[1]{\href{https://orcid.org/#1}{\mbox{\scalerel*{
\begin{tikzpicture}[yscale=-1,transform shape]
\pic{orcidlogo};
\end{tikzpicture}
}{|}}}}
\newcommand*{\field}[1]{\mathbb{#1}}

\begin{document}
\title{Discrete-Time Modeling of Interturn Short Circuits in Interior PMSMs}

\author{
	\vskip 1em
	Lukas Zezula \orcidicon{0000-0002-3183-2438}, Matus Kozovsky \orcidicon{0000-0002-1547-1003}, Ludek Buchta \orcidicon{0000-0002-8954-3495} and Petr Blaha \orcidicon{0000-0001-5534-2065}
	
	\thanks{
		This work was supported in part by European Union through the project Robotics and Advanced Industrial Production under Grant CZ.02.01.01/00/22\_008/0004590, in part by the Ministry of Education, Youth and Sports of the Czech Republic and Chips Joint Undertaking through the project Archimedes: Trusted lifetime in operation for a circular economy No 101112295/9A23010, in part by Czech Science Foundation through the Analysis of Discrete and Continuous Dynamical Systems with Emphasis on Identification Problems Grant 23-06476S, and in part by the infrastructure of RICAIP that has received funding from the European Union's Horizon 2020 research and innovation programme under grant agreement No 857306 and from Ministry of Education, Youth and Sports under OP RDE grant agreement No CZ.02.1.01/0.0/0.0/17\_043/0010085. \emph{(Corresponding author: Lukas Zezula.)}
		
		The authors are with the CEITEC - Central European Institute of Technology, Brno University of Technology, 612 00 Brno, Czech Republic (e-mail: lukas.zezula@ceitec.vutbr.cz).
		
		This work has been submitted to the IEEE for possible publication. Copyright may be transferred without notice, after which this version may no longer be accessible.
	}
}

\maketitle
	
\begin{abstract}
This article describes the discrete-time modeling approach for interturn short circuits in interior permanent magnet synchronous motors with concentrated windings that facilitate model-based fault diagnostics and mitigation. A continuous-time model incorporating universal series-parallel stator winding connection and radial permanent magnet fluxes is developed in the stator variables and transformed into the rotor reference frame, including also the electromagnetic torque. The transformed model undergoes discretization using the matrix exponential-based technique, wherein the electrical angular velocity and angle are considered time-varying parameters. The resulting model is subsequently expanded to consider the motor connection resistance via perturbation techniques. In the laboratory experiments, we validate the dynamical properties of the derived model by comparing its outputs with the experimental data and waveforms generated by the forward Euler-based discrete-time approximation.
\end{abstract}


\markboth{}%
{}

\definecolor{limegreen}{rgb}{0.2, 0.8, 0.2}
\definecolor{forestgreen}{rgb}{0.13, 0.55, 0.13}
\definecolor{greenhtml}{rgb}{0.0, 0.5, 0.0}

\section{Introduction}
\IEEEPARstart{P}{resently}, at the forefront of electromobility, the ongoing research of autonomous electric vehicles (EVs) has increased the safety and availability requirements for electric powertrains, impacting even interior permanent magnet synchronous motors (PMSMs), which are preferred over surface-mounted PMSMs in EV propulsions, especially due to their superior efficiency and wide speed range. Consequently, as the traditional approach of transferring a motor to a safe state upon fault detection lacks both the safety and the availability, modern electric drives must incorporate fail operational (FO) strategies \cite{Failop1}, \cite{Failop2}, thus being able to endure one internal failure and continue operating at their standard or with reduced capabilities. Robust controller design techniques can then be integrated to achieve passive insensitiveness to different faults; however, this often produces reduced system ratings in a failure-free performance \cite{FTC}. Thus, modern FO drives should be based on active controller reconfiguration after fault detection and localization. Model predictive control (MPC) fault mitigation algorithms \cite{Fault_mit1}, \cite{Fault_mit2}, \cite{Fault_mit3}, \cite{Fault_mit4}, \cite{Fault_mit5}, \cite{Fault_mit6} then embody a promising avenue for suppressing current harmonic distortions and torque ripple in a post-fault operation while ensuring minimum power loss.

The effectiveness of the MPC-based fault mitigation methods depends heavily on the precision of a discrete-time model describing a shorted motor and the accuracy of the parameters utilized. Although discrete time is fundamental in MPC for employing numerical techniques to optimize the control actions in each prediction step, most modeling approaches focus exclusively on continuous-time descriptions \cite{Mod_simp1}, \cite{Mod_simp2}, \cite{Mod_simp3} and the subsequent precision enhancement by reflecting advanced winding architectures \cite{Mod_seg1}, \cite{Mod_seg2}, \cite{Mod_seg3}, \cite{Mod_seg4} and other motor nonidealities (e.g., non-sinusoidal fluxes \cite{Mod_nonideal1}, \cite{Mod_nonideal2} or asymmetric windings \cite{Mod_nonideal3}) in the model. Discrete-time models (DTMs) are then predominantly derived using the forward Euler technique \cite{Fault_mit3}, \cite{Fault_mit4}, \cite{Fault_mit5}, \cite{Fault_mit6}, \cite{Disc_Eul_Ctrl1}, \cite{Disc_Eul_Ctrl2}, \cite{Disc_Eul_diag1}, \cite{Disc_Eul_diag2}, resulting in more straightforward descriptions when compared to approaches employing higher-order Taylor series DTM approximations \cite{Disc_Taylor}. Conversely, as the dynamics of PMSMs are velocity-dependent, the method involves significant prediction errors and numerical instability at higher rates \cite{Disc}. Moreover, an interturn short circuit (ISC) leads to an imbalance of the stator windings' impedance that is accompanied by trigonometric nonlinearities in a shorted motor description. The rapid oscillations with steep gradients then further aggravate the errors of the forward Euler approximations. As the model error accumulates with each prediction step, employing the forward Euler DTMs is problematic especially in monitoring the electrical parameters or estimating the fault indicators where long-term statistics are observed \cite{Fault_diag}.

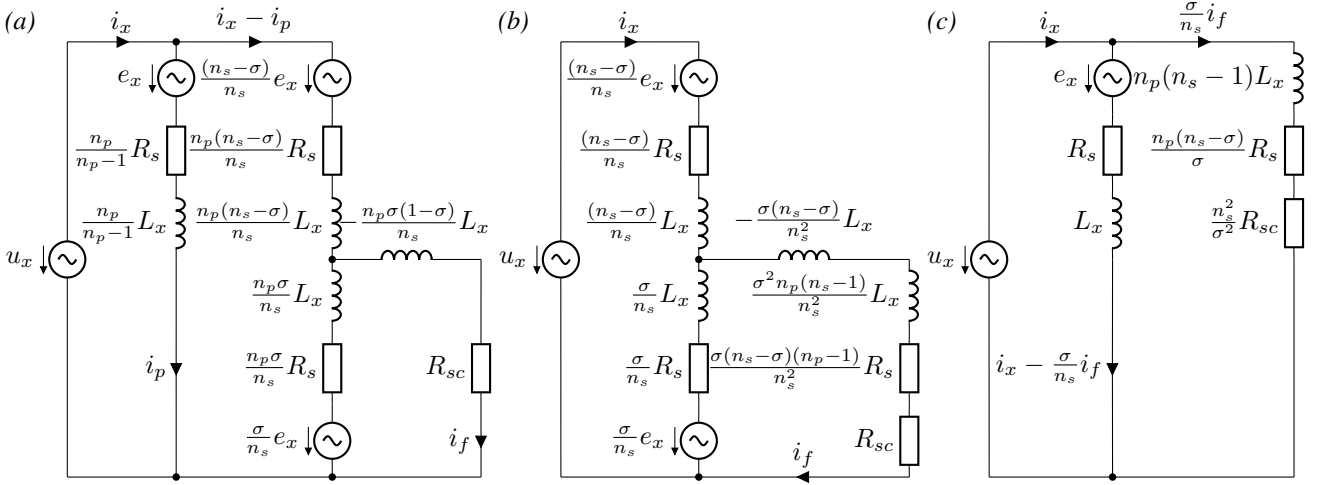
\begin{figure*}[t]
\begin{center}
 \begin{circuitikz}[scale=0.965]
 \ctikzset{resistor = european}
 \ctikzset{inductor = american}
 \def\cone{1.5}
 \def\ctwo{3.65}
 \def\cthree{5.7}
 \def\cfour{6.8}
 \def\cfive{8.7}
 \def\csix{11.6}
\def\cseven{12.7}
 \def\ceight{14.4}
  \def\cnine{16.9}
 \node[] at (-0.65,3.27) {\textit{(a)}};
  \draw (0,3) to[/tikz/circuitikz/bipoles/length=0.8cm,vsourcesin, v_=$u_{x}$] (0,-3);
  \draw[black] (0, 3) to[short, i=$i_x$] (\cone, 3);
  \node[circle, fill=black, inner sep=1pt] at (\cone, 3){};
  \draw (\cone,3) to[/tikz/circuitikz/bipoles/length=0.8cm,vsourcesin, v_=$e_x$] (\cone, 2);
  \draw (\cone,2) to[/tikz/circuitikz/bipoles/length=0.8cm,R, l_=$\frac{n_p}{n_p-1}R_s$] (\cone, 1);
  \draw (\cone,1) to[/tikz/circuitikz/bipoles/length=0.8cm,L, l_=$\frac{n_p}{n_p-1}L_x$] (\cone, 0);
  \draw[black] (\cone, 0) to[short, i_=$i_p$] (\cone, -3);
  \node[circle, fill=black, inner sep=1pt] at (\cone, -3){};
  \draw[black] (0, -3) -- (\cone, -3);
  \draw[black] (\cone, 3) to[short, i=$i_x-i_p$] (\ctwo, 3);
  \draw (\ctwo,3) to[/tikz/circuitikz/bipoles/length=0.8cm,vsourcesin, v_=$\frac{(n_s-\sigma)}{n_s}e_x$] (\ctwo, 2);
  \draw (\ctwo,2) to[/tikz/circuitikz/bipoles/length=0.8cm,R, l_=$\frac{n_p(n_s-\sigma)}{n_s}R_s$] (\ctwo, 1);
  \draw (\ctwo,1) to[/tikz/circuitikz/bipoles/length=0.8cm,L, l_=$\frac{n_p(n_s-\sigma)}{n_s}L_x$] (\ctwo, 0);
  \draw (\ctwo,0) to[/tikz/circuitikz/bipoles/length=0.8cm,L, l_=$\frac{n_p\sigma}{n_s}L_x$] (\ctwo, -1);
  \draw (\ctwo,-1) to[/tikz/circuitikz/bipoles/length=0.8cm,R, l_=$\frac{n_p\sigma}{n_s}R_s$] (\ctwo, -2);
  \draw (\ctwo,-2) to[/tikz/circuitikz/bipoles/length=0.8cm,vsourcesin, v_=$\frac{\sigma}{n_s} e_x$] (\ctwo, -3);
  \node[circle, fill=black, inner sep=1pt] at (\ctwo, -3){};
  \draw[black] (\cone, -3) -- (\ctwo, -3);
  \node[circle, fill=black, inner sep=1pt] at (\ctwo, 0){};
  \draw (\ctwo, 0) to[/tikz/circuitikz/bipoles/length=0.8cm,L,name=L1] (\cthree, 0);
  \node[above] at (L1.n) {\hspace{2mm}$-\frac{n_p\sigma(1-\sigma)}{n_s}L_x$};
  \draw (\cthree,0) to[/tikz/circuitikz/bipoles/length=0.8cm,R, l_=$R_{sc}$] (\cthree, -3);
  \draw[black] (\cthree, -2.4999) to[short, i_=$i_f$] (\cthree, -2.5001);
  \draw[black] (\cthree, -3) -- (\ctwo, -3);
 \node[] at (6.15,3.27) {\textit{(b)}};
  \draw (\cfour,3) to[/tikz/circuitikz/bipoles/length=0.8cm,vsourcesin, v_=$u_{x}$] (\cfour,-3); 
  \draw[black] (\cfour, 3) to[short, i=$i_x$] (\cfive, 3);
  \draw (\cfive,3) to[/tikz/circuitikz/bipoles/length=0.8cm,vsourcesin, v_=$\frac{(n_s-\sigma)}{n_s}e_x$] (\cfive, 2);
  \draw (\cfive,2) to[/tikz/circuitikz/bipoles/length=0.8cm,R, l_=$\frac{(n_s-\sigma)}{n_s}R_s$] (\cfive, 1);
  \draw (\cfive,1) to[/tikz/circuitikz/bipoles/length=0.8cm,L, l_=$\frac{(n_s-\sigma)}{n_s}L_x$] (\cfive, 0);
  \draw (\cfive,0) to[/tikz/circuitikz/bipoles/length=0.8cm,L, l_=$\frac{\sigma}{n_s}L_x$] (\cfive, -1);
  \draw (\cfive,-1) to[/tikz/circuitikz/bipoles/length=0.8cm,R, l_=$\frac{\sigma}{n_s}R_s$] (\cfive, -2);
  \draw (\cfive,-2) to[/tikz/circuitikz/bipoles/length=0.8cm,vsourcesin, v_=$\frac{\sigma}{n_s} e_x$] (\cfive, -3);
  \draw[black] (\cfour, -3) -- (\cfive, -3);
  \node[circle, fill=black, inner sep=1pt] at (\cfive, 0){};
  \draw (\cfive, 0) to[/tikz/circuitikz/bipoles/length=0.8cm,L, name=L2] (\csix, 0);
  \node[above] at (L2.n) {$-\frac{\sigma(n_s-\sigma)}{n_s^2}L_x$};
  \draw (\csix,0) to[/tikz/circuitikz/bipoles/length=0.8cm,L, l_=$\hspace{2mm}\frac{\sigma^2n_p(n_s-1)}{n_s^2}L_x$] (\csix, -1);
  \draw (\csix,-1) to[/tikz/circuitikz/bipoles/length=0.8cm,R, l_=$\frac{\sigma(n_s-\sigma)(n_p-1)}{n_s^2}R_s$] (\csix, -2);
  \draw (\csix,-2) to[/tikz/circuitikz/bipoles/length=0.8cm,R, l_=$R_{sc}$] (\csix, -3);
  \draw[black] (\cfive/2+\csix/2+0.00001, -3) to[short, i_=$i_f$] (\cfive/2+\csix/2-0.00001, -3);
  \draw[black] (\cfive, -3) -- (\csix, -3);
  \node[circle, fill=black, inner sep=1pt] at (\cfive, -3){};
 \node[] at (12.05,3.27) {\textit{(c)}};
  \draw (\cseven,3) to[/tikz/circuitikz/bipoles/length=0.8cm,vsourcesin, v_=$u_{x}$] (\cseven,-3); 
  \draw[black] (\cseven, 3) to[short, i=$i_x$] (\ceight, 3);
  \draw (\ceight,3) to[/tikz/circuitikz/bipoles/length=0.8cm,vsourcesin, v_=$e_x$] (\ceight, 2);
  \draw (\ceight,2) to[/tikz/circuitikz/bipoles/length=0.8cm,R, l_=$R_s$] (\ceight, 1);
  \draw (\ceight,1) to[/tikz/circuitikz/bipoles/length=0.8cm,L, l_=$L_x$] (\ceight, 0);
  \draw[black] (\ceight, 0) to[short, i_=$i_x - \frac{\sigma}{n_s}i_f$] (\ceight, -3);
  \draw[black] (\cseven, -3) -- (\ceight, -3);
\draw[black] (\ceight, 3) to[short, i=$\frac{\sigma}{n_s}i_f$] (\cnine, 3);
  \node[circle, fill=black, inner sep=1pt] at (\ceight, 3){};
    \draw (\cnine,3) to[/tikz/circuitikz/bipoles/length=0.8cm,L, l_=$n_p(n_s-1)L_x$] (\cnine, 2);
    \draw (\cnine,2) to[/tikz/circuitikz/bipoles/length=0.8cm,R, l_=$\frac{n_p(n_s-\sigma)}{\sigma}R_s$] (\cnine, 1);
    \draw (\cnine,1) to[/tikz/circuitikz/bipoles/length=0.8cm,R, l_=$\frac{n_s^2}{\sigma^2}R_{sc}$] (\cnine, 0);
    \draw[black] (\cnine, 0) -- (\cnine, -3);
  \node[circle, fill=black, inner sep=1pt] at (\ceight, -3){};
  \draw[black] (\ceight, -3) -- (\cnine, -3);
 \end{circuitikz}
 \end{center}
 \vspace{-0.4cm}
\caption{The shorted phase $x$ equivalent circuits: \textit{(a)} full; \textit{(b)} algebraically reduced; \textit{(c)} with reduced currents directly related to $u_x$.}\vspace{-0.25cm}
\label{fig:eq_circ}
\end{figure*}

This article aims to derive an advanced discrete-time model of interior PMSMs with an ISC in a concentrated winding segment applicable to model-based fault diagnostics and mitigation. The model considers series-parallel stator winding connection and radial distribution of permanent magnet fluxes. The discretization is performed by the matrix exponential-based approach, which has proved efficient in \cite{Disc}, \cite{Fault_diag}, and \cite{Fault_curr_mon}. Afterwards, a motor connection resistance is added to the DTM using perturbation techniques.

Generally, this article is organized as follows: Section \ref{sec:cont_mod} focuses on the continuous-time modeling of ISCs in interior PMSMs; Section \ref{sec:disc_mod} discusses the discretization of the continuous-time description; Section \ref{sec:modexp} proposes DTM extension via perturbation methods; Section \ref{sec:valid} validates the derived DTM; and, finally, Section \ref{sec:conclusion} concludes the article.

\section{Continous-time model}
\label{sec:cont_mod}
In FO PMSMs, the stator windings are typically composed of series-parallel connected, geometrically separated coil segments. The separation effectively mitigates the probability of ISC spreading across an entire winding, yet mutual inductive couplings exist even between the distinct segments in the stator phase. However, once an ISC has occurred in one of the segments, the couplings between the concentrated winding parts become negligible compared to the mutual inductance in the fault-affected segment. A model comprising independent coils with uniform inductance $\frac{n_p}{n_s}L_x$ and resistance $\frac{n_p}{n_s}R_s$ can then represent the stator winding. The winding arrangement-related parameters $n_p$ and $n_s$ refer to a connection where $n_p$ parallel branches contain $n_s$ coils in series, and $R_s$ and $L_x$ denote the overall resistance and inductance attributable to a stator phase in the absence of an ISC. The diagrams in Fig. \ref{fig:eq_circ} show various equivalent circuits representing an arbitrary shorted stator phase $x\in\{a,b,c\}$. In Fig. \ref{fig:eq_circ}, $u_{x}$ stands for the stator phase $x$ voltage, $e_x$ signifies the voltage induced by the permanent magnets and other phases, and $i_x$, $i_p$, and $i_f$ denote the currents through the entire phase $x$, parallel branches lacking an ISC, and an ISC, respectively. The fault-related parameters $R_{sc}$ and $\sigma\in\langle0,1\rangle$ then refer to the short circuit resistance and the winding segment portion that is shorted by $R_{sc}$. Using mesh current analysis on the equivalent circuit in Fig. \ref{fig:eq_circ} \textit{(a)}, an algebraic dependency of $i_p$ on $i_x$ and $i_f$ is derived. We have
\begin{equation}
i_p = \frac{n_p-1}{n_p} \left(i_x - \frac{\sigma}{n_s}i_f\right).
\label{eqn:ip}
\end{equation}
Hence, as shown in Fig. \ref{fig:eq_circ} \textit{(b)} and \textit{(c)}, the equivalent circuits of the shorted phase $x$ can be modeled using only $i_x$ and $i_f$.

Suppose a wye-connected interior PMSM with an ISC in the phase $a$. Due to the wye connection, the stator currents $\textbf{i}_{abc} = \begin{bmatrix} i_{a} & i_{b} & i_{c} \end{bmatrix}^T$ must satisfy $i_a+i_b+i_c=0$, but the fault current present in one of the phases $\textbf{i}_{f} = \begin{bmatrix} i_f & 0 & 0 \end{bmatrix}^T$ leads to an imbalance in the voltage equations. Consequently, according to the equivalent circuit in Fig.~\ref{fig:eq_circ} \textit{(c)}, the stator voltages $\textbf{u}_{abc} = \begin{bmatrix} u_{a} & u_{b} & u_{c} \end{bmatrix}^T$ read
\begin{align}
\textbf{u}_{abc} &= \textbf{R}\,\textbf{i}_{abc}- \frac{\sigma}{n_s}\textbf{R}\,\textbf{i}_{f} + \diff{}{t}\left(\textbf{L}\,\textbf{i}_{abc}- \frac{\sigma}{n_s}\textbf{L}\,\textbf{i}_{f} + \Lambda_{pm}\right)\nonumber\\
\Lambda_{pm} &= \sum_{j=1,3,5,\ldots} \lambda_{pm,j} \begin{bmatrix}
\cos\Bigl(j\theta_e + \phi_{\lambda,j}\Bigl) \\ \cos\Bigl(j\left(\theta_e-\frac{2\pi}{3}\right)+ \phi_{\lambda,j}\Bigl) \\ \cos\Bigl(j\left(\theta_e+\frac{2\pi}{3}\right)+ \phi_{\lambda,j}\Bigl)
\end{bmatrix}.
\label{eqn:fault_sys_abc}
\end{align}
In (\ref{eqn:fault_sys_abc}), $\lambda_{pm,j}$ and $\phi_{\lambda,j}$ stand for the amplitudes and phase shifts towards $\phi_{\lambda,1}=0$ of the radial permanent magnet flux linkage, $\textbf{R}^{3\times 3} = \text{diag}\left(R_s, R_s, R_s\right)$ represent the diagonal resistance matrix, and $\theta_e$ is the electrical angle. Due to the symmetry of a motor, the inductive couplings between the phases have equal strengths, and the symmetric inductance matrix $\textbf{L}^{3\times 3} = \begin{bmatrix} L_{ij}\end{bmatrix}$ is used in the model. The elements in $\textbf{L}$ are defined as in
\begin{align}
L_{ij} &= \begin{cases}
L_s + L_{fl}\cos\left(2\theta_e\right) & ij \in \{aa\}\\
L_s + L_{fl}\cos\left(2\theta_e + \frac{2\pi}{3}\right) & ij \in \{bb\}\\
L_s + L_{fl}\cos\left(2\theta_e - \frac{2\pi}{3}\right) & ij \in \{cc\}\\
-L_m + L_{fl}\cos\left(2\theta_e\right) & ij \in \{bc,cb\}\\
-L_m + L_{fl}\cos\left(2\theta_e + \frac{2\pi}{3}\right) & ij \in \{ac,ca\}\\
-L_m + L_{fl}\cos\left(2\theta_e - \frac{2\pi}{3}\right) & ij \in \{ab,ba\}\\
\end{cases}
\label{eqn:induct_self}
\end{align}
where $L_s$ and $L_m$ are the phase self and mutual inductances, and $L_{fl}$ represents the inductances' fluctuation with the changing rotor angle. The inductances defined in the stator variables (\textit{abc}) are equivalently represented in the rotor reference frame (\textit{dq}), as follows:
\begin{align}
	L_d &= L_s + L_m + 1.5L_{fl}&&
	L_q = L_s + L_m - 1.5L_{fl}\nonumber\\
	L_0 &= L_s - 2 L_m. &&
	\label{eqn:induct_dq}
\end{align}
In (\ref{eqn:induct_dq}), $L_d$, $L_q$, and $L_0$ refer to the direct, quadrature, and zero-sequence inductance, respectively.

The stator voltages $\textbf{u}_{abc}$, introduced in model (\ref{eqn:fault_sys_abc}), are determined by the difference between the voltage potentials at the motor terminals $\textbf{u}_{t} = \begin{bmatrix}u_{a,t} & u_{b,t} & u_{c,t}\end{bmatrix}^T$ and the central point of the wye-connected phases $u_0$. The potentials $\textbf{u}_{t}$ are then traditionally defined as in
\begin{align}
	\textbf{u}_{t} &= 
	\begin{bmatrix} 
		\cos\left(\theta_e\right) & -\sin\left(\theta_e\right)\\ 
		\cos\left(\theta_e - \frac{2\pi}{3}\right) & -\sin\left(\theta_e - \frac{2\pi}{3}\right)\\ 
		\cos\left(\theta_e + \frac{2\pi}{3}\right) & -\sin\left(\theta_e + \frac{2\pi}{3}\right)\\ 
	 \end{bmatrix}\begin{bmatrix}
	 u_d \\ u_q
	 \end{bmatrix} + \begin{bmatrix}
	 u_m \\ u_m \\ u_m
	 \end{bmatrix}
	\label{eqn:volt_terminal}
\end{align}
where the dominant part of $\textbf{u}_{t}$ is obtained by transforming the voltage control actions $u_d$ and $u_q$ from the rotor reference frame to \textit{abc}, and $u_m$ represents the modulation signal introduced to use all the available DC bus voltage effectively. The central point potential $u_0 = -(u_a+u_b+u_c)/3 + u_m$ can then be derived from the stator voltages definition (\ref{eqn:fault_sys_abc}), as follows:
\begin{align}
u_{0} &= u_m + \frac{1}{3}\frac{\sigma}{n_s}R_si_f + \frac{1}{3}\frac{\sigma}{n_s}L_0\diff{i_f}{t} - \diff{\lambda_{pm}^0}{t}\nonumber\\
\lambda_{pm}^0 &=  \sum_{j=3,9,15,\ldots} \lambda_{pm,j}\cos\Bigl(j\theta_e + \phi_{\lambda,j}\Bigl).
\label{eqn:zero_volt}
\end{align}
Subsequently, by substituting for the stator voltages in (\ref{eqn:fault_sys_abc}), we obtain
\begin{align}
\textbf{u}_{t}-\left(u_m-\diff{\lambda_{pm}^0}{t}\right)\begin{bmatrix} 1 \\ 1 \\ 1\end{bmatrix} &= \textbf{R}\,\textbf{i}_{abc,h} + \diff{}{t}\left(\textbf{L}\,\textbf{i}_{abc,h} + \Lambda_{pm}\right)
\label{eqn:fault_sys_abc2}
\end{align}
where $\textbf{i}_{abc,h} = \begin{bmatrix} i_{a,h} & i_{b,h} & i_{c,h}\end{bmatrix}^T$ are the healthy parts of a shorted motor's stator currents $\textbf{i}_{abc}$, defined as in
\begin{align}
\begin{bmatrix}
i_{a,h} \\ i_{b,h} \\ i_{c,h}
\end{bmatrix}=  \begin{bmatrix}
i_{a} \\ i_{b} \\ i_{c}
\end{bmatrix} + \frac{1}{3}\frac{\sigma}{n_s}\begin{bmatrix}
-2i_f \\ i_f \\ i_f
\end{bmatrix}.
\label{eqn:curr_output}
\end{align}
Note that (\ref{eqn:fault_sys_abc2}) corresponds to the model of a healthy interior PMSM, and currents $\textbf{i}_{abc,h}$ are balanced, $i_{a,h}+i_{b,h}+i_{c,h}=0$, enabling the \textit{dq} transformation without the zero-sequence component. Analogously, the differential equation governing $i_f$ is derived from the equivalent circuit in Fig. \ref{fig:eq_circ} \textit{(c)} using $u_a = u_{a,t} - u_0$ and definition (\ref{eqn:zero_volt}), as follows:
\begin{align}
u_{a,t} - u_m &= R_f i_f + \diff{}{t} \left(L_f i_f - \lambda_{pm}^0 \right) \nonumber\\
R_f &= n_p\left(1-\frac{\sigma}{n_s}\right)R_s+\frac{1}{3}\frac{\sigma}{n_s}R_s+\frac{n_s}{\sigma}R_{sc}\nonumber\\
L_f &= \frac{\sigma}{n_s}n_p(n_s-1)L_{aa}+\frac{1}{3}\frac{\sigma}{n_s}L_0.
\label{eqn:fault_curr_abc}
\end{align}
Finally, an electromagnetic torque $T_e$ can be obtained through an analysis of the total energy supplied by a motor, $W_e = W_l + W_f$, as in
\begin{align}
	W_e &= \int \textbf{i}_{abc}^T\textbf{u}_{abc}\text{d}t = \int \textbf{i}_{abc,h}^T\textbf{u}_{t} + \frac{\sigma}{n_s}i_f (u_{a,t}-u_m)\text{d}t\nonumber\\
	W_l &= \int \textbf{i}_{abc,h}^T \textbf{R} \textbf{i}_{abc,h} + \frac{\sigma}{n_s} R_f i_f^2 \text{d}t\nonumber\\
	W_f &= \frac{1}{2} \textbf{i}_{abc,h}^T \textbf{L} \textbf{i}_{abc,h} + \frac{1}{2} \frac{\sigma}{n_s} L_f i_f^2 + \textbf{i}_{abc,h}^T \Lambda_{pm} - \frac{\sigma}{n_s}i_f\lambda_{pm}^0\nonumber\\
	T_e &= \frac{\partial W_f}{\partial \theta_m} = P_P \frac{\partial W_f}{\partial \theta_e}
	\label{eqn:torque_derivation}
\end{align}
where $W_l$ is the heat loss, $W_f$ denote the energy stored in the coupling field, $\theta_m$ stands for the mechanical angle, and $P_P$ represents the number of pole pairs. A similar approach can then be employed to model ISCs in the remaining subsystem phases, with the differences occurring in (\ref{eqn:curr_output}), where an ISC in a different stator phase results in a permutation of the fault current contributions, and in (\ref{eqn:fault_curr_abc}), where $L_{aa}$ and $u_{a,t}$ are substituted with either $L_{bb}$ and $u_{b,t}$ or $L_{cc}$ and $u_{c,t}$.

As (\ref{eqn:fault_sys_abc2}) corresponds to the healthy model of an interior PMSM, it can be equivalently represented in the rotor reference frame. We have
\begin{align}
u_{d} &= R_s i_{d,h} + L_d\diff{i_{d,h}}{t} -\omega_e\left(L_q i_{q,h} + \lambda_{pm}^q\right) \nonumber\\
u_{q} &= R_s i_{q,h} + L_q\diff{i_{q,h}}{t} +\omega_e\left(L_d i_{d,h} + \lambda_{pm}^d\right)
\label{eqn:fault_sys_dq}
\end{align}
where currents $\textbf{i}_{abc,h}$ are transformed into $i_{d,h}$ and $i_{q,h}$; $\omega_e$ represents the electrical angular velocity; and the contributions of radial permanent magnet fluxes $\lambda_{pm}^d$ and $\lambda_{pm}^q$ read
\begin{align}
	\lambda_{pm}^d =&\sum_{j=6,12,18,\ldots} \begin{bmatrix}(1-j)\lambda_{pm,j-1}\\ (j+1)\lambda_{pm,j+1}\end{bmatrix}^T\begin{bmatrix}\cos(j\theta_e+\phi_{\lambda,j-1}) \\ \cos(j\theta_e+\phi_{\lambda,j+1})\end{bmatrix}\nonumber\\  &+\lambda_{pm,1}\nonumber\\ 
	\lambda_{pm}^q =& \sum_{j=6,12,18,\ldots} \begin{bmatrix}(j-1)\lambda_{pm,j-1}\\ (j+1)\lambda_{pm,j+1}\end{bmatrix}^T\begin{bmatrix}\sin(j\theta_e+\phi_{\lambda,j-1}) \\ \sin(j\theta_e+\phi_{\lambda,j+1})\end{bmatrix}.
	\label{eqn:pm_fluxes_dq}
\end{align}
The fault current is then analogously described using the \textit{dq} voltages and inductances, as follows:
\begin{align}
\diff{L_f i_f}{t} &= -R_f i_f + \begin{bmatrix}u_{d}\\u_{q}\end{bmatrix}^T \begin{bmatrix}\cos\left(\theta_e+\phi_f\right)\\-\sin\left(\theta_e+\phi_f\right) \end{bmatrix} + \omega_e \frac{\partial \lambda_{pm}^0}{\partial \theta_e}\nonumber\\
L_f &= L_{f1} + L_{f2} \cos\left(2\theta_e-\phi_f\right)\nonumber\\
L_{f1} &= \frac{\sigma}{n_s}n_p(n_s-1)\frac{L_d+L_q+L_0}{3} + \frac{1}{3}\frac{\sigma}{n_s}L_0\nonumber\\
L_{f2} &= \frac{\sigma}{n_s}n_p(n_s-1)\frac{L_d-L_q}{3}
\label{eqn:fault_curr_dq}
\end{align}
where $R_f$ was established in (\ref{eqn:fault_curr_abc}), and $\phi_f = \{0,-2\pi/3,2\pi/3\}$ if the ISC is in phase $\{a,b,c\}$, respectively. Similarly, the coupling equation (\ref{eqn:curr_output}) is transformed as in
\begin{align}
\begin{bmatrix}
i_{d} \\ i_{q}
\end{bmatrix} =  \begin{bmatrix}
i_{d,h} \\ i_{q,h}
\end{bmatrix} + \frac{2}{3}\frac{\sigma}{n_s}i_f\begin{bmatrix}
\cos\left(\theta_e+\phi_f\right)\\-\sin\left(\theta_e+\phi_f\right)
\end{bmatrix}.
\label{eqn:curr_output_dq}
\end{align}
As shown above, the \textit{dq} currents involve the healthy model outputs and fault current contributions. The same principle applies even for $T_e$. We have
\begin{align}
T_e =&  \frac{3}{2}P_P(\lambda_{pm}^di_{q,h}-\lambda_{pm}^qi_{d,h} + (L_d - L_q)i_{d,h}i_{q,h}) \nonumber\\
&- P_P \frac{\sigma}{n_s}L_{f2}i_f^2\sin(2\theta_e-\phi_f) - P_P \frac{\sigma}{n_s}i_f \frac{\partial \lambda_{pm}^0}{\partial \theta_e}\nonumber\\
\frac{\partial \lambda_{pm}^0}{\partial \theta_e} =& -\sum_{j=3,9,15,\ldots} j\lambda_{pm,j}\sin\Bigl(j\theta_e + \phi_{\lambda,j}\Bigl).
\label{eqn:torque}
\end{align}
As seen in (\ref{eqn:torque}), the zero-sequence flux $\lambda_{pm}^0$ not affecting the healthy machine performance influences, in a combination with $i_f$, the torque ripple in post-fault machine operation.

\section{Discrete-time model}
\label{sec:disc_mod}
Due to the higher inertia and limited maximum electromagnetic torque, the angular velocity maintains an almost constant value within the sampling interval, producing a piecewise linear angle. We have
\begin{align}
\omega_e &=\omega_e(t) \approx \omega_e(k) \nonumber\\
\theta_e &= \theta_e(t) \approx \theta_e(k) + (t-kT_s)\omega_e(k)
\label{eqn:disc_vel_angl}
\end{align}
where $T_s$ stands for the sampling period, and $k$ denotes the current DTM step. Approximations (\ref{eqn:disc_vel_angl})  remain valid even after an ISC has occurred, as the integral relationship between velocity and torque, along with the higher inertia, effectively suppresses the harmonic distortion pattern in (\ref{eqn:torque}).

The DTM of a healthy interior PMSM with sinusoidal permanent magnet fluxes was derived in \cite{Disc}. Assuming the radial fluxes, the DTM of (\ref{eqn:fault_sys_dq}) reads
\begin{align}
\begin{bmatrix}
i_{d,h}(k+1)\\i_{q,h}(k+1)
\end{bmatrix} &= e^{\textbf{A}T_s}\begin{bmatrix}
i_{d,h}(k)\\i_{q,h}(k)
\end{bmatrix} + \int_{0}^{T_s}e^{\textbf{A}t}\textbf{L}_{dq}^{-1}\begin{bmatrix}
u_d\\u_q
\end{bmatrix} \text{d}t + \textbf{Q}_k \nonumber\\
\textbf{Q}_k &= \omega_e(k) \int_{0}^{T_s}e^{\textbf{A}t}\textbf{L}_{dq}^{-1}\begin{bmatrix}
	 \lambda_{pm}^q\\- \lambda_{pm}^d
\end{bmatrix} \text{d}t
\label{eqn:disc_eqv_3ph}
\end{align}
where the $u_d$, $u_q$, $\lambda_{pm}^d$, and $\lambda_{pm}^q$ occurring in the integrals above are evaluated at time $(k+1)T_s-t$, $\textbf{L}_{dq} = \text{diag}(L_d,L_q)$, and $\textbf{A}=\textbf{A}(t=kT_s)$ expresses the system matrix of the \textit{dq} model state-space representation. The step-dependent exponential of the system matrix $\textbf{E}_k = e^{\textbf{A}T_s}$ was approximated in \cite{Disc}, as follows:
\begin{align}
\textbf{E}_k &\approx e^{-\rho T_s} \textbf{L}_{dq}^{-1} \textbf{T}(T_s) \textbf{L}_{dq} + e^{-\rho T_s} \frac{\sin(\omega_e T_s)}{\omega_e} \text{diag}(\delta,-\delta) \nonumber\\
\textbf{T}(T_s) &= \begin{bmatrix} \cos\left(\omega_e T_s\right) & \sin\left(\omega_e T_s\right)\\ -\sin\left(\omega_e T_s\right) & \cos\left(\omega_e T_s\right) \end{bmatrix}
\label{eqn:mat_exp}
\end{align}
where $\omega_e = \omega_e(k)$, and substitutions $\rho$ and $\delta$ are defined as in
\begin{align}
\rho =  \frac{R_s(L_d+L_q)}{2L_dL_q} && \delta =  \frac{R_s(L_d-L_q)}{2L_dL_q}.
\label{eqn:subs}
\end{align}
In \cite{Disc}, the solution to the integral in (\ref{eqn:disc_eqv_3ph}) involved assuming constant $u_d$ and $u_q$ over the sampling period. However, since the voltages connected to the motor terminals are produced by inverter switching and the stator currents are acquired when the distortion resulting from the pulse width modulation is minimal, the assumption of constant \textit{dq} voltages leads to biased current estimations because the effect of a changing electrical angle in a sampling interval is omitted. This problem can be solved by considering constant motor terminal potentials $\textbf{u}_t$ in a sampling period rather than constant $u_d$ and $u_q$, yielding
\begin{align}
\begin{bmatrix}
u_d\bigl((k+1)T_s-t\bigl) \\ u_q\bigl((k+1)T_s-t\bigl)
\end{bmatrix} = \textbf{T}^{-1}(t) \textbf{T}(T_s)\begin{bmatrix}
u_d(k) \\ u_q(k)
\end{bmatrix}
\label{eqn:input_disc}
\end{align}
where the rotation matrix, $\textbf{T}(t)$, was defined in (\ref{eqn:mat_exp}) as a parameter-dependent function. Note that $\textbf{T}(t)$ commutes $\textbf{T}^{-1}(t)\textbf{T}(T_s)=\textbf{T}(T_s)\textbf{T}^{-1}(t)$, and its inverse corresponds to transposition $\textbf{T}^{-1}(t)=\textbf{T}^{T}(t)$. Subsequently, the discrete-time model (\ref{eqn:disc_eqv_3ph}) can be written as in
\begin{align}
	\begin{bmatrix}
		i_{d,h}(k+1)\\i_{q,h}(k+1)
	\end{bmatrix} = \textbf{E}_k\begin{bmatrix}
		i_{d,h}(k)\\i_{q,h}(k)
	\end{bmatrix} + \textbf{B}_k\begin{bmatrix}
		u_d(k)\\u_q(k)
	\end{bmatrix} + \textbf{Q}_{k}
	\label{eqn:disc_eqv_3ph_2}
\end{align}
where $\textbf{B}_k$ is obtained by solving the integral
\begin{align}
\textbf{B}_k &= \int_{0}^{T_s}e^{\textbf{A}t}\textbf{L}_{dq}^{-1} \textbf{T}^{-1}(t) \textbf{T}(T_s) \text{d}t.
\label{eqn:integrals_sys_mat}
\end{align}
Evaluating $\textbf{B}_k$ and $\textbf{Q}_k$ then requires integrating the exponentially damped sine and cosine functions. Although an analytical solution is available, we used approximations to avoid the rational functions that involve $\omega_e$ in the DTM. The approximations read
\begin{align}
	I_1 = \int_0^{T_s} e^{-\rho t}\sin(n \omega_e t) \text{d}t &\approx \frac{1-\cos(n \omega_e T_s)}{n \omega_e}e^{-\frac{\rho T_s}{2}} = \hat{I}_1\nonumber\\
	I_2 = \int_0^{T_s} e^{-\rho t}\cos(n \omega_e t) \text{d}t &\approx \frac{\sin(n \omega_e T_s)}{n \omega_e}e^{-\frac{\rho T_s}{2}}= \hat{I}_2
	\label{eqn:integrals_aprox}
\end{align}
where the parameter $n\in\field{N}$. We computed the errors of \eqref{eqn:integrals_aprox} on the intervals of feasible values $0<\rho\leq1/T_s$ and $|\omega_e|\leq2\pi/T_s$, as in $|I_1-\hat{I}_1|\leq 0.1T_s/n$ and $|I_2-\hat{I}_2|\leq 0.026T_s$. Using (\ref{eqn:integrals_aprox}), we derived $\textbf{B}_k$ and $\textbf{Q}_{k}$ (\ref{eqn:integrals_solved}). 

A similar discretization with approximations (\ref{eqn:disc_vel_angl}) then applies even to the fault current description (\ref{eqn:fault_curr_dq}), giving the DTM
\begin{align}
	i_f(k+1) &= \frac{a_{f}(T_s)L_{f,k}}{L_{f,k+1}}i_f(k)+\frac{\textbf{b}_{f,k}^T}{L_{f,k+1}}\begin{bmatrix}u_{d}(k)\\u_{q}(k)\end{bmatrix}+\frac{q_{f,k}}{L_{f,k+1}} \nonumber\\
	L_{f,k} &= L_{f1} + L_{f2} \cos\bigl(2\theta_e(k)-\phi_f\bigl)\nonumber\\
	L_{f,k+1} &= L_{f1} + L_{f2} \cos\bigl(2\theta_e(k)+2\omega_eT_s-\phi_f\bigl)\nonumber\\
	a_{f}(\tau) &= e^{-\int_0^{\tau}R_f/\left(L_{f1} + L_{f2} \cos\bigl(2\theta_e(k)+2\omega_e(T_s-t)-\phi_f\bigl)\right)\text{d}t} \nonumber\\
	\textbf{b}_{f,k} &= \int_0^{T_s}a_{f}(\tau)\text{d}\tau\begin{bmatrix}\cos\left(\theta_e(k)+\phi_f\right)\\-\sin\left(\theta_e(k)+\phi_f\right) \end{bmatrix} \nonumber\\
	q_{f,k} &= -\omega_e\int_0^{T_s}a_{f}(\tau)\sum_{j=3,9,15,\ldots} j\lambda_{pm,j}\sin\bigl(\theta_{3}(\tau)\bigl) \text{d}\tau\nonumber\\
	\theta_{3}(\tau) &= j\theta_e(k)+j\omega_e(T_s-\tau) + \phi_{\lambda,j}.
	\label{eqn:fault_curr_disc}\tag{25}
\end{align}

\begin{strip}\vspace{-0.65cm}\hrule\vspace{-0.1cm}
	\begin{align}
		\textbf{Q}_k &\approx e^{-\frac{\rho T_s}{2}}\textbf{L}_{dq}^{-1}\left(-\lambda_{pm,1}\begin{bmatrix}
			1-\cos(\omega_e T_s)\\ \sin(\omega_e T_s) - \delta \frac{1-\cos(\omega_e T_s)}{\omega_e}\end{bmatrix}+\sum_{j=6,12,18,\ldots} \lambda_{pm,j-1}\textbf{M}_{1}\begin{bmatrix}
			\cos(\theta_{1}) \\ -\sin(\theta_{1})
		\end{bmatrix} + \lambda_{pm,j+1}\textbf{M}_{2}\begin{bmatrix}
			\cos(\theta_{2}) \\ \sin(\theta_{2})
		\end{bmatrix} \right)\nonumber\\
		\textbf{M}_{1} &= \textbf{T}(T_s) - \textbf{T}(j T_s) + \frac{\sin(\omega_e T_s)}{\omega_e} \text{diag}(\delta,-\delta) + \text{diag}(\delta,-\delta)\frac{\textbf{T}^{-1}(T_s)-\textbf{T}(jT_s)}{(j+1)\omega_e}\begin{bmatrix} 0& -1\\ 1& 0\end{bmatrix} \hspace{1.755cm} \theta_1 = j \theta_e(k) + \phi_{\lambda,j-1}\nonumber\\
		\textbf{M}_{2} &= \textbf{T}(T_s) - \textbf{T}^{-1}(j T_s) + \frac{\sin(\omega_e T_s)}{\omega_e} \text{diag}(\delta,-\delta) + \text{diag}(\delta,-\delta)\frac{\textbf{T}^{-1}(T_s)-\textbf{T}^{-1}(jT_s)}{(j-1)\omega_e}\begin{bmatrix} 0& 1\\ -1& 0\end{bmatrix} \hspace{1cm} \theta_2 = j \theta_e(k) + \phi_{\lambda,j+1}\nonumber\\
		\textbf{B}_k &\approx \textbf{L}_{dq}^{-1}\frac{1-e^{-\rho T_s}}{\rho}\textbf{T}(T_s) - \textbf{L}_{dq}^{-1}\frac{\sin(\omega_e T_s)}{\omega_e} \frac{e^{-\rho T_s}-e^{-\frac{\rho T_s}{2}}}{\rho}\text{diag}(\delta,-\delta)
		\label{eqn:integrals_solved}\tag{24}\setcounter{equation}{25}
	\end{align}
\end{strip}

\noindent Even though coefficient $a_{f}(\tau)$ can be expressed analytically, its subsequent integration is intractable; therefore, we utilized its first-order Taylor series approximation close to point $L_{f2}\to0$. We have
\begin{align}
	a_{f}(\tau) &\approx  e^{-\frac{R_f}{L_{f1}}\tau}+\frac{L_{f2}}{L_{f1}}\frac{R_f}{L_{f1}}e^{-\frac{R_f}{L_{f1}}\tau} h_f(\tau)\nonumber\\
	h_f(\tau) &= \begin{bmatrix}\frac{1-\cos(2\tau\omega_e)}{2\omega_e} \\ \frac{\sin(2\tau\omega_e)}{2\omega_e}\end{bmatrix}^T\textbf{T}(2T_s) \begin{bmatrix}\sin\bigl(2\theta_e(k)-\phi_f\bigl) \\ \cos\bigl(2\theta_e(k)-\phi_f\bigl)\end{bmatrix}
	\label{eqn:coef_a}
\end{align}
where $L_{f1}$ and $L_{f2}$ were defined in (\ref{eqn:fault_curr_dq}), and functional $\textbf{T}(\ldots)$ corresponds to (\ref{eqn:mat_exp}). Employing approximation (\ref{eqn:coef_a}), the quotients $\textbf{b}_{f,k}$ and $q_{f,k}$ are expressed as in (\ref{eqn:fault_coef}).
\begin{figure*}[b]\vspace{-0.5cm}\hrule
	\begin{align}
		\textbf{b}_{f,k} &\approx \left(\frac{L_{f1}}{R_f}\left(1-e^{-\frac{R_f}{L_{f1}}T_s}\right)-\frac{L_{f2}}{L_{f1}}\left(e^{-\frac{R_f}{L_{f1}}T_s}-e^{-\frac{R_f}{2L_{f1}}T_s}\right)h_f(T_s)\right)\begin{bmatrix}\cos\left(\theta_e(k)+\phi_f\right)\\-\sin\left(\theta_e(k)+\phi_f\right) \end{bmatrix} \nonumber\\
		q_{f,k} &\approx -e^{-\frac{R_f}{2L_{f1}}T_s}\sum_{j=3,9,15,\ldots} \lambda_{pm,j} \begin{bmatrix}\cos\bigl(j\theta_e(k)+\phi_{\lambda,j}\bigl)\\ \sin\bigl(j\theta_e(k)+\phi_{\lambda,j}\bigl) \end{bmatrix}^T\left(\begin{bmatrix} 1-\cos(jT_s\omega_e) \\ \sin(jT_s\omega_e)\end{bmatrix} + \frac{1}{2}\frac{L_{f2}}{L_{f1}}\frac{R_f}{L_{f1}}\textbf{M}_3\begin{bmatrix}\sin\bigl(2\theta_e(k)-\phi_{f}\bigl)\\ \cos\bigl(2\theta_e(k)-\phi_{f}\bigl)\end{bmatrix}\right) \nonumber\\
		\textbf{M}_3 &= \begin{bmatrix}-2 & 0\\ 0 & 0\end{bmatrix}\frac{\textbf{I} - \textbf{T}(2T_s)}{2\omega_e} + \begin{bmatrix}-1&0\\0&1\end{bmatrix}\frac{\textbf{I} - \textbf{T}^{-1}\bigl((j-2)T_s\bigl)}{(j-2)\omega_e}+\frac{\textbf{I} - \textbf{T}\bigl((j+2)T_s\bigl)}{(j+2)\omega_e} \hspace{2cm} \textbf{I} = \begin{bmatrix}1&0\\0&1\end{bmatrix}
		\label{eqn:fault_coef}
	\end{align}
\end{figure*}

The derived DTM then comprises difference equations (\ref{eqn:disc_eqv_3ph_2}) and (\ref{eqn:fault_curr_disc}) (with time-varying quotients (\ref{eqn:mat_exp}), (\ref{eqn:integrals_solved}), (\ref{eqn:coef_a}), and (\ref{eqn:fault_coef})) and algebraic relations (\ref{eqn:curr_output_dq}) and (\ref{eqn:torque}) describing the output currents and electromagnetic torque, evaluated at time $kT_s$. Given that $\sin(\omega_e T_s)/\omega_e$, $\bigl(1-\cos(\omega_e T_s)\bigl)/\omega_e$, and $\bigl(\cos(\omega_e T_s)-\cos(j \omega_e T_s)\bigl)/\bigl((j\pm1)\omega_e\bigl)$ are always less than $T_s$, their corresponding terms can be ignored to further simplify the quotients. The velocity $\omega_e$ and angle $\theta_e$ in the DTM are treated as input variables.

\section{Perturbation-based model extension}
\label{sec:modexp}
The derived DTM provides a basis for further model advancement using, for example, perturbation techniques \cite{Perturb_method}. For instance, the model can be refined to consider the resistance of the motor connection $R_{c}$, comprising the dynamic resistances of the power stage components (switches, diodes), connectors, cables, and terminal box elements. Generally, $R_c$ causes a voltage drop of $R_{c}i_x$, where $i_x$ is the stator current in phase $x$, reflectable in the \textit{dq} model as follows:
\begin{align}
u_{d} =& (R_s+R_c) i_{d,h} + L_d\diff{i_{d,h}}{t} -\omega_e\left(L_q i_{q,h} + \lambda_{pm}^q\right) \nonumber\\
       &+ \frac{2}{3}\frac{\sigma}{n_s}R_c i_f\cos\left(\theta_e+\phi_f\right)\nonumber\\
u_{q} =& (R_s+R_c) i_{q,h} + L_q\diff{i_{q,h}}{t} +\omega_e\left(L_d i_{d,h} + \lambda_{pm}^d\right)\nonumber\\
&- \frac{2}{3}\frac{\sigma}{n_s}R_c i_f\sin\left(\theta_e+\phi_f\right) \nonumber\\
\diff{L_f i_f}{t} =& -\left(R_f+\frac{2}{3}\frac{\sigma}{n_s}R_c\right) i_f + \begin{bmatrix}u_{d}\\u_{q}\end{bmatrix}^T \begin{bmatrix}\cos\left(\theta_e+\phi_f\right)\\-\sin\left(\theta_e+\phi_f\right) \end{bmatrix}\nonumber\\
&-R_c\begin{bmatrix}i_{d,h}\\i_{q,h}\end{bmatrix}^T \begin{bmatrix}\cos\left(\theta_e+\phi_f\right)\\-\sin\left(\theta_e+\phi_f\right) \end{bmatrix}+ \omega_e \frac{\partial \lambda_{pm}^0}{\partial \theta_e}.
\label{eqn:fault_sys_dq_Rc}
\end{align}
As shown in (\ref{eqn:fault_sys_dq_Rc}), the presence of $R_c$ increases the resistance within the model and produces couplings between the healthy model part and the fault current description. These couplings then drastically increase the analytical discretization complexity. Conversely, ignoring the couplings yields a model structurally equivalent to that in Section \ref{sec:cont_mod}, the only distinction being increased resistances. The discrete-time equivalent of model (\ref{eqn:fault_sys_dq_Rc}) can then be structured with the dominant parts following the DTM presented in Section \ref{sec:disc_mod}, where the resistances in the healthy model and the fault current equation are replaced with $R_s+R_c$ and $R_f+\frac{2}{3}\frac{\sigma}{n_s}R_c$, respectively, and the perturbated parts capturing the problematic couplings. In such a case, as derived in \cite{Perturb_method}, the state-transition matrix is approximable as in
\begin{align}
\Phi&=\Phi\bigl((k+1)T_s,kT_s\bigl)\approx\begin{bmatrix}\textbf{E}_k & \Delta_h v^T\\ v\Delta_f & \frac{a_{f}(T_s)L_{f,k}}{L_{f,k+1}} \end{bmatrix} \nonumber\\
v &= \begin{bmatrix}\cos\bigl(\theta_e(k)+\phi_f\bigl) & -\sin\bigl(\theta_e(k)+\phi_f\bigl)\end{bmatrix}
\label{eqn:ST_matrix}
\end{align}
where the main diagonal contains the initially computed transition terms, while the perturbed couplings are positioned in the antidiagonal. Matrices $\Delta_h$ and $\Delta_f$ read
\begin{align}
\Delta_h &= -\frac{2}{3}\frac{\sigma}{n_s}R_c\int_0^{T_s} e^{\textbf{A}t}\frac{a_f(T_s)}{a_f(t)}\frac{L_{f,k}}{L_{f,t}}\textbf{L}_{dq}^{-1}\textbf{T}(T_s)\textbf{T}^{-1}(t)\text{d}t\nonumber\\
L_{f,t} &= L_{f1} + L_{f2} \cos\bigl(2\theta_e(k)+2\omega_e(T_s-t)-\phi_f\bigl)\nonumber\\
\Delta_f &= -\frac{R_c}{L_{f,k+1}}\int_0^{T_s}a_f(t)\textbf{T}(t)\textbf{T}^{-1}(T_s)e^{\textbf{A}(T_s-t)}\text{d}t.
\label{eqn:cross_coupl}
\end{align}
 The integrals in (\ref{eqn:cross_coupl}) were approximated using the first-order Taylor series close to the point $L_d-L_q\to0$, yielding
\begin{align}
\Delta_h &\approx -\frac{2}{3}\frac{\sigma}{n_s}R_c\frac{e^{-\frac{R_f^*}{L_{f1}}T_s}-e^{-\rho^* T_s}\left(2-\frac{L_{f,k}}{L_{f1}}\right)}{\rho^*-\frac{R_f^*}{L_{f1}}}\textbf{L}_{dq}^{-1}\textbf{T}(T_s)\nonumber\\
\Delta_f &\approx -\frac{L_d+L_q}{2}\frac{R_c}{L_{f,k+1}}\frac{e^{-\frac{R_f^*}{L_{f1}}T_s}-e^{-\rho^* T_s}}{\rho^*-\frac{R_f^*}{L_{f1}}}\text{diag}\left(\frac{1}{L_q},\frac{1}{L_d}\right)
\label{eqn:cross_coupl2}
\end{align}
where $R_f^*=R_f+\frac{2}{3}\frac{\sigma}{n_s}R_c$ and $\rho^*=(R_s+R_c)(L_d+L_q)/(2L_dL_q)$. In (\ref{eqn:cross_coupl2}), the terms weighted with a number lower than $T_s$ were ignored to preserve tolerable complexity. The presence of terms in the antidiagonal of the state-transition matrix (\ref{eqn:ST_matrix}) influences both the state variables and the model inputs. However, the input contribution can be ignored as it scales with the second power of $T_s$. The discrete-time approximation of (\ref{eqn:fault_sys_dq_Rc}) is then computed as in
\begin{align}
\begin{bmatrix}
	i_{d,h}(k+1)\\i_{q,h}(k+1)\\i_f(k+1)
	\end{bmatrix} = \Phi\begin{bmatrix}
		i_{d,h}(k)\\i_{q,h}(k)\\i_f(k)
	\end{bmatrix} +\begin{bmatrix}\textbf{B}_k \\ \frac{\textbf{b}_{f,k}^T}{L_{f,k+1}}\end{bmatrix} + \begin{bmatrix}
		\textbf{Q}_{k} \\[5pt] \frac{q_{f,k}}{L_{f,k+1}}\\ 
	\end{bmatrix}.
\label{eqn:DTM_Rc}
\end{align}

\section{Validation}
\label{sec:valid}

\begin{figure}[t!]
	\centering
        \def\svgwidth{\columnwidth}
	\input{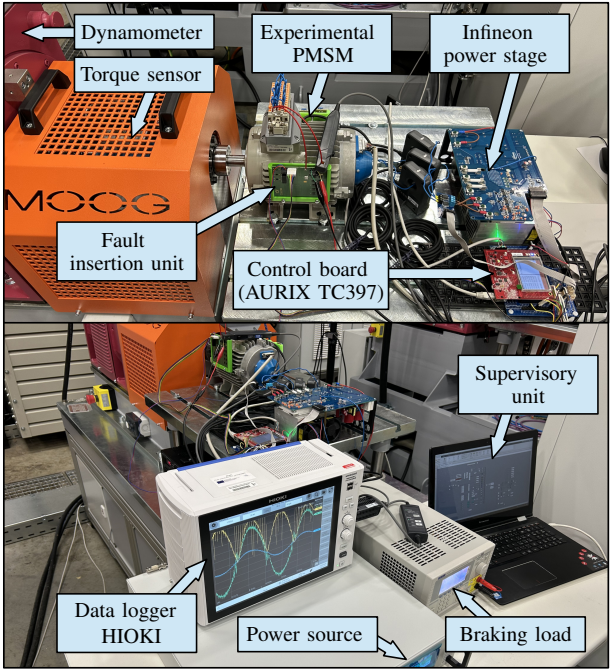}
        \vspace{-1cm}  
	\caption{The testbed with the experimental motor.}\vspace{-0.25cm}
	\label{fig:wb}
\end{figure}

We validated the proposed DTM by comparing its outputs with the measurements on the experimental motor and the data generated by the DTM derived using the forward Euler discretization approach applied to (\ref{eqn:fault_sys_dq_Rc}). The motor is driven by an EVAL-M1-IR2214 power stage, controlled via a KIT\_A2G\_TC397 application board featuring an AURIX TC397 microcontroller with $T_s=100$~\unit{\micro\second}. Since the power stage lacks suitable current sensing, additional Hall effect-based sensors, MCA1101-20-5, are used for the phase current measurement. The motor is mechanically coupled to a dynamometer, which supports the torque and speed operation modes. Moreover, the dynamometer incorporates a high-precision Kistler 4503B torque sensor with a high-frequency digital output, enabling an accurate measurement of the torque generated by the motor on the shaft. An image of the testbed with the experimental motor is displayed in Fig. \ref{fig:wb}. 
\begin{figure}[t!]
	\centering
    \def\svgwidth{\columnwidth-1cm}
	\input{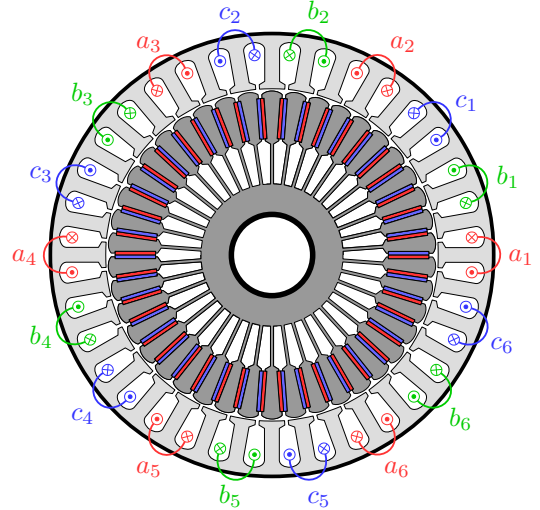}
    \vspace{-0.7cm}  
	\caption{The internal structure of the experimental motor.}\vspace{-0.2cm}
	\label{fig:MotStructure}
\end{figure}

\subsection{The experimental motor structure and parameters}
\label{sec:mot_struct}
In the validation experiments, we utilized an interior PMSM, with its internal structure shown in Fig. \ref{fig:MotStructure}. The stator phase contains six winding segments, each including 25 turns, individually connected to the terminal box to allow various series-parallel stator winding configurations. However, the validation was performed using only series-connected coils. Unlike industrial motors, where the phase segment-linking resistance is negligible, the experimental PMSM features the resistance of two copper wires (each 17 cm in length, with a cross-section of 1 mm$^2$) linking the coils to the ILME CDSHM 42 terminal box, as well as the transient resistance of its connectors, which is rated as less than 3 \unit{\milli\ohm}. Thus, the segment linking resistance can reach up to 11.8~\unit{\milli\ohm}, 9.7\% of the 121.2~\unit{\milli\ohm} stator segment resistance average (the connections excluded) measured with a HIOKI LCR HiTESTER 3511-50. While the nonnegligible segment linking resistance integrates smoothly into the series-connected model through an increase of $R_c$, it violates (\ref{eqn:ip}) when $n_p>1$, introducing an additional differential equation. Therefore, due to the experimental motor's distinctive structure, we opted for the connection with $n_p=1$ and $n_s=6$.
\begin{table}[t]
	\caption{The experimental motor parameters}\vspace{-0.4cm}
	\label{tab:params}
    \setlength\tabcolsep{3pt}
    \def\arraystretch{1.1}
    \begin{center}
	\begin{tabular}{|p{1.3cm}|p{0.85cm}||p{1.3cm}|p{0.85cm}||p{1.75cm}|p{0.85cm}|}
		\hline
		Parameter & Value & Parameter & Value & Parameter & Value \\
		\hline
        $R_s$ [\unit{\milli\ohm}] & 727 &$L_0$ [\unit{\milli\henry}] & 2.74 & $\lambda_{pm,1}$ [\unit{\milli\weber}] & 18.4\\
        \hline
        $R_c$ [\unit{\milli\ohm}] & 362 & $P_P$ [-] & 21 & $\phi_{\lambda,1}$ [\unit{\degree}] & 0 \\
        \hline
        $L_d$ [\unit{\milli\henry}] & 3.29 & $n_p$ [-] & 1 & $\lambda_{pm,3}$ [\unit{\micro\weber}] & 200 \\
        \hline
        $L_q$ [\unit{\milli\henry}] & 3.12 & $n_s$ [-] & 6 & $\phi_{\lambda,3}$ [\unit{\degree}] & 0\\
        \hline 
	\end{tabular}\vspace{-0.4cm}
    \end{center}
\end{table}
\begin{figure*}[t!]
	\centering
    \input{Build/Experiment_layout}
    \vspace{-0.3cm}
    \caption{The layout of the experiments.}\vspace{-0.25cm}
	\label{fig:explayout}
\end{figure*}

The parameters of the experimental motor were identified and listed in Table \ref{tab:params}. Firstly, we measured $R_s$ and $L_s$ using the LCR meter and extracted the dominant permanent magnet flux components via the fast Fourier transform from the back electromotive force measured with a HIOKI MR6000-01 data logger at multiple velocities controlled by the dynamometer. Subsequently, $R_s+R_c$, $L_d$, and $L_q$ were estimated at different current levels and averaged through the locked-rotor test with pseudo-random binary sequence voltage excitation. Finally, the estimated values of $L_d$ and $L_q$, along with the measured $L_s$, were used to compute $L_0$ as in (\ref{eqn:induct_dq}).

\subsection{Fault insertion unit}
\label{sec:FIU}
The experimental motor features winding taps connected to the terminal box that facilitate emulating different ISC severities. Physical faults can then be emulated by electrically connecting specific winding taps or via a specially designed fault insertion unit (FIU) that allows controlling digitally its resistance $R_{FIU}$. The FIU comprises multiple back-to-back SiR180DP power MOSFETs operating in the linear region. The transistors are controlled by an analog circuit incorporating four stages of operational amplifiers. In the first stage, differential amplifiers measure the voltage across the 0.25 \unit{\milli\ohm} shunt resistor $u_{shunt}$ and the FIU terminal $u_{FIU}$. The voltage $u_{shunt}$ is subsequently amplified in the second stage with four digital inputs governing the switching between the pre-defined gain levels. The third stage computes the absolute values of $u_{FIU}$ and the multiplied $u_{shunt}$, and the final one compares these using an operational amplifier and adjusts the transistors' gate voltage to maintain balance. This configuration keeps the ratio between $u_{FIU}$ and $u_{shunt}$ on the gain selected in the second stage, allowing digital control of $R_{FIU}$. 

In the validation experiments, the applied gain configuration resulted in $R_{FIU}=\{442, 47.0, 5.62, 1.74\}$~\unit{\milli\ohm}, with $R_{FIU}=1.74$ \unit{\milli\ohm} corresponding to the fully open MOSFETs state. The FIU resistance values were calculated from the FIU voltage and current waveforms acquired by the data logger during the fault emulation experiments. The total short-circuit resistance $R_{sc}=R_{FIU}+R_{wire}$ then includes also $R_{wire}=14.4$~\unit{\milli\ohm}, comprising the resistance of the FIU-to-winding taps cables and connectors. Emulating faults by means of additional wiring introduces even the parasitic inductance $L_{wire}$, which influences $i_f$ through increases in $L_{f1}$. We have
\begin{align}
L_{f1} &= \frac{\sigma}{n_s}n_p(n_s-1)\frac{L_d+L_q+L_0}{3} + \frac{1}{3}\frac{\sigma}{n_s}L_0 + \frac{n_s}{\sigma}L_{wire}
\label{eqn:wire_ind}
\end{align}
where we identified $L_{wire} = 3.81$~\unit{\micro\henry}. As shown in (\ref{eqn:wire_ind}), the inductance increase follows the reciprocal value of $\sigma$, making lower numbers of shorted turns more affected.

\subsection{Model implementation}
\label{sec:implementation}
The derived DTM was integrated into the AURIX TC397 microcontroller, running concurrently with the control algorithm and the forward Euler DTM. As illustrated in Fig. \ref{fig:explayout}, the models receive inputs from the control algorithm, and their outputs are transmitted together with the measured data to the supervisory unit, which provides a velocity setpoint for the dynamometer in the speed mode and the torque request for the experimental motor. As shown in Fig. \ref{fig:explayout}, the control algorithm implements the maximum torque per ampere (MTPA), field weakening (FW), maximum torque per volt (MTPV), space vector modulation, and dead time (DT) compensation, while $\omega_e$ is estimated using an angle tracking observer (ATO). As also indicated in Fig. \ref{fig:explayout}, we implemented a specially designed firmware in the AURIX microcontroller to process the torque sensor digital output based on edge counting and precise timestamping. The torque and velocity data, zero-phase digital-filtered by the low-pass second-order Butterworth filter (cutoff at 500 \unit{rad/\second}), are subsequently employed to determine the electromagnetic torque based on the identified friction and known inertias in the supervisory unit.

We measured the computational times of the implemented models on the AURIX TC397 and, additionally, on an STM32G474RE to assess the performance across the different computational platforms. We opted for the STM32G474RE because it embodies a low-cost microcontroller with mainstream relevance. The results are presented in Table \ref{tab:CompTime}. While the derived DTM requires additional computations compared to the forward Euler DTM, primarily due to the evaluation of $\sin(T_s\omega_e)$ and $\cos(T_s\omega_e)$, its processing time remains negligible considering the microcontrollers' capabilities, making it computationally efficient.
\begin{table}[t]
	\vspace{-0.3cm}
	\caption{The computational time of the implemented DTMs in \unit{\micro\second}}\vspace{-0.4cm}
	\label{tab:CompTime}
    \setlength\tabcolsep{2pt}
    \def\arraystretch{1.1}
	\begin{center}
		\begin{tabular}{|c|l|l|l|l|l|}
			\hline
			\begin{tabular}[c]{@{}c@{}}MCU\end{tabular}                    & \multicolumn{1}{c|}{\begin{tabular}[c]{@{}c@{}}Measured\\ Time \end{tabular}} &
			\multicolumn{1}{c|}{\begin{tabular}[c]{@{}c@{}}Pre-fault \\Euler DTM\end{tabular}} & \multicolumn{1}{c|}{\begin{tabular}[c]{@{}c@{}}Post-fault \\Euler DTM\end{tabular}} &
			\multicolumn{1}{c|}{\begin{tabular}[c]{@{}c@{}}Pre-fault \\derived DTM \end{tabular}} &
			\multicolumn{1}{c|}{\begin{tabular}[c]{@{}c@{}}Post-fault \\derived DTM \end{tabular}} \\  \hline  
			\multirow{3}{*}{\begin{tabular}[c]{@{}c@{}}STM32\\G474\end{tabular}}  
            & Min      & 2.06   & 5.01     & 4.95   & 12.95    \\ \cline{2-6} 
			& Average  & 2.06   & 5.01   & 4.95   & 12.95   \\ \cline{2-6} 
			& Max      & 2.39   & 5.32   & 5.30   & 13.33   \\ \hline
			\multirow{3}{*}{\begin{tabular}[c]{@{}c@{}}TC397\end{tabular}} 
            & Min      & 0.67   & 1.53   & 1.35  & 3.73   \\ \cline{2-6} 
			& Average  & 0.75   & 1.57   & 1.39  & 3.77   \\ \cline{2-6} 
			& Max      & 0.92   & 1.77   & 1.60   & 3.95   \\ \hline
		\end{tabular}\vspace{-0.4cm}
	\end{center}	
\end{table}

\begin{figure*}[t!]
	\centering
	\includegraphics[]{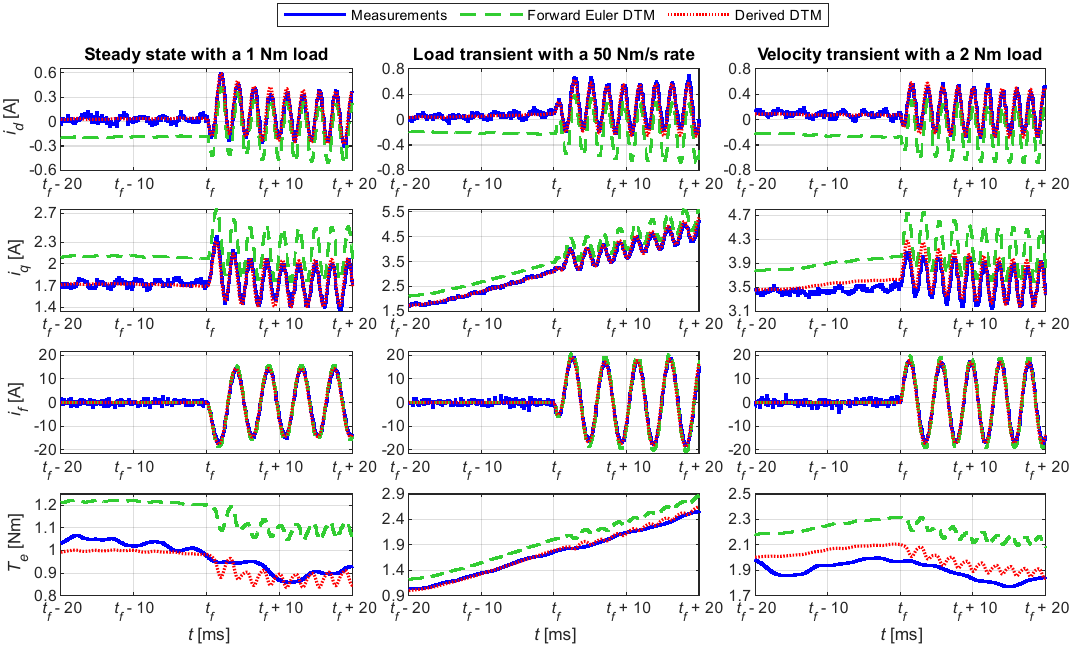}\vspace{-0.3cm}
	\caption{The model-to-measurement comparison across the operating conditions.}
	\label{fig:OP_Cond}
\end{figure*}
\begin{figure*}[t!]
	\centering
	\includegraphics[]{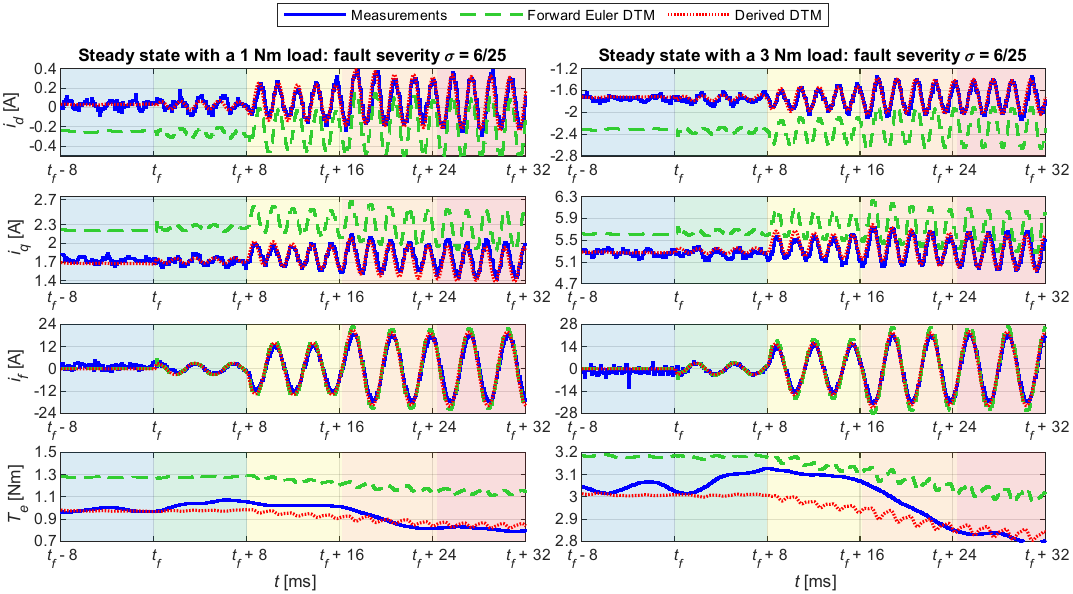}
 	\includegraphics[]{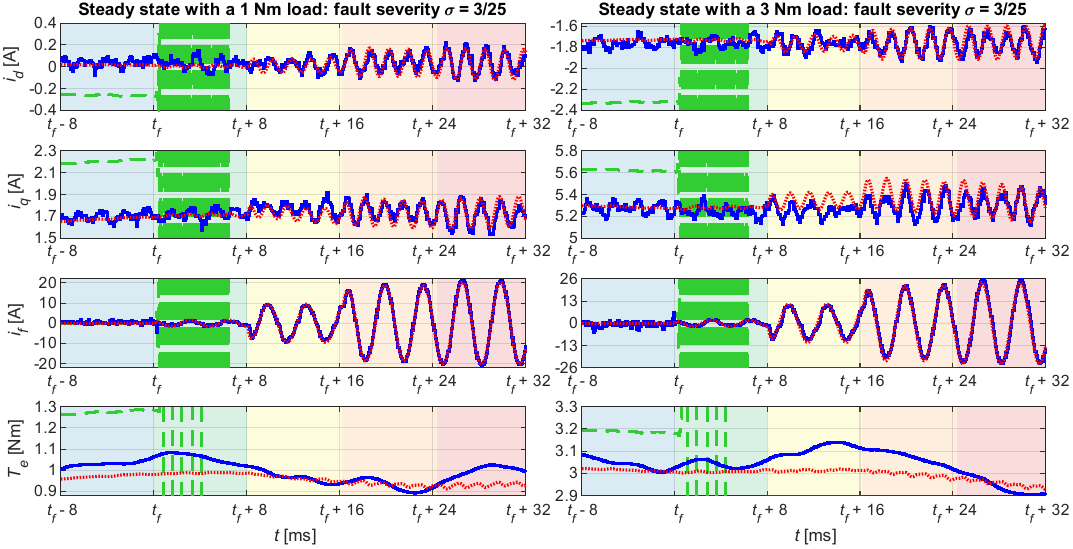}\vspace{-0.3cm}
	\caption{The model-to-measurement comparison across the fault severities and ISC resistances.}
	\label{fig:Rsc_changes}
\end{figure*}

\subsection{Experiments}
\label{sec:experiments}
One part of the validation experiments focused on the DTMs' predictions across different operating conditions, namely, the velocity steady state at 1400~\unit{rad/\second} with a load of 1~\unit{Nm}, load transient at 1400~\unit{rad/\second} from 1 to 3~\unit{Nm} with a 50~\unit{Nm/\second} rate, and velocity transient from 1200 to 1600~\unit{rad/\second} with a 10000~\unit{rad/\second}$^2$ rate and 2~\unit{Nm} load. The ISCs were emulated at time $t_f$ with the fault severity of $10/25$ and $R_{FIU} = 1.74$ \unit{\milli\ohm}. The outcomes of the experiments across the diverse operating conditions are shown in Fig.~\ref{fig:OP_Cond}. As indicated in Fig.~\ref{fig:OP_Cond}, the forward Euler DTM's predictions exhibit an offset because this model inherently assumes constant \textit{dq} voltages over the sampling period, which does not align with the actual behavior. Furthermore, $i_f$ predicted by this model is overrated compared to the measured data. Conversely, the derived DTM predictions closely match the measured waveforms, except for the velocity transients, where slightly higher errors occur. These errors stem from the ATO’s inability to track rapidly changing velocity, producing a biased angular velocity estimation provided to both models. The absence of harmonic distortion in the measured $T_e$ is due to $T_e$ being computed via filtered signals during post-processing.

The other part of the validation experiments tested the models under the fault severities of $6/25$ and $3/25$, with $1$ and $3$~\unit{Nm} torque loads and real-time FIU resistance changes, decreasing within $R_{FIU}=\{442, 47.0, 5.62, 1.74\}$~\unit{\milli\ohm}. The experiments were conducted at the $\omega_e = 1900$~\unit{rad/\second} steady state, with the fault occurring at $t_f$. The model-to-measurement comparison across the fault severities and ISC resistances is presented in Fig. \ref{fig:Rsc_changes} using color-coded resistance changes. As shown in Fig. \ref{fig:Rsc_changes}, aside from the biased responses and overrated fault current predicted by the forward Euler DTM at $\sigma=6/25$, this model even becomes numerically unstable at $\sigma=3/25$ after the fault occurrence. The instability arises from violating the Euler method's condition: the time constants must exceed the sampling period. Hence, the forward Euler DTM cannot simulate an ISC in its early stages, characterized by minor $\sigma$ with higher $R_{sc}$, namely, low time constant $L_{f1}/R_f^*$. Conversely, the derived DTM is numerically stable across the feasible domain and tracks the actual behavior. The better fit of $i_f$ in Fig.~\ref{fig:OP_Cond} compared to Fig.~\ref{fig:Rsc_changes} is linked to a lower sensitivity to $L_{wire}$ at higher $\sigma$. As $L_{wire}$ cannot be reliably measured, it was only roughly estimated, ignoring that the actual $L_{wire}$ varies with the emulated $\sigma$ due to the diverse positioning of the FIU-to-winding taps cables in the motor case.

\section{Conclusion}
\label{sec:conclusion}
 \begin{table}[t]\vspace{-0.2cm}
	\caption{Comparing the ISC modeling approaches}\vspace{-0.2cm}
	\label{tab:comparison}
	\definecolor{mygreen}{RGB}{50, 205, 50}
	\setlength\tabcolsep{0.005cm}
	\begin{tabular}{@{}>{\centering\arraybackslash}m{3.3cm}|| >{\centering\arraybackslash}m{1.1cm} >{\centering\arraybackslash}m{0.6cm} >{\centering\arraybackslash}m{0.6cm} >{\centering\arraybackslash}m{0.6cm} >{\centering\arraybackslash}m{0.6cm} >{\centering\arraybackslash}m{0.6cm} >{\centering\arraybackslash}m{0.6cm} >{\centering\arraybackslash}m{0.6cm}@{}}
		Model features & Derived model &\cite{Fault_mit4}&\cite{Mod_simp1}&\cite{Mod_seg1}&\cite{Mod_seg2}&\cite{Mod_seg4}&\cite{Mod_nonideal1}&\cite{Disc_Eul_diag2}\\
		\hline\hline
		
		Radial permanent magnet fluxes
		&\hspace{0.07cm}{\color{mygreen}\checkmark}&\hspace{0.04cm}{\color{red}$\times$}&\hspace{0.07cm}{\color{mygreen}\checkmark}&\hspace{0.04cm}{\color{red}$\times$}&\hspace{0.04cm}{\color{red}$\times$}&\hspace{0.04cm}{\color{red}$\times$}&\hspace{0.07cm}{\color{mygreen}\checkmark}&\hspace{0.04cm}{\color{red}$\times$}\\  	
		
		Series/parallel winding connection &\hspace{0.07cm}{\color{mygreen}\checkmark}&\hspace{0.04cm}{\color{red}$\times$}&\hspace{0.04cm}{\color{red}$\times$}&\hspace{0.07cm}{\color{mygreen}\checkmark}&\hspace{0.07cm}{\color{mygreen}\checkmark}&\hspace{0.07cm}{\color{mygreen}\checkmark}&\hspace{0.04cm}{\color{red}$\times$}&\hspace{0.04cm}{\color{red}$\times$}\\  	
		
		Inductive couplings between the stator phase segments&\hspace{0.04cm}{\color{red}$\times$}&\hspace{0.04cm}{--}&\hspace{0.04cm}{--}&\hspace{0.07cm}{\color{mygreen}\checkmark}&\hspace{0.07cm}{\color{mygreen}\checkmark}&\hspace{0.07cm}{\color{mygreen}\checkmark}&\hspace{0.04cm}{--}&\hspace{0.04cm}{--}\\

        Connection resistance&\hspace{0.07cm}{\color{mygreen}\checkmark}&\hspace{0.04cm}{\color{red}$\times$}&\hspace{0.04cm}{\color{red}$\times$}&\hspace{0.04cm}{\color{red}$\times$}&\hspace{0.04cm}{\color{red}$\times$}&\hspace{0.04cm}{\color{red}$\times$}&\hspace{0.07cm}{\color{mygreen}\checkmark}& \hspace{0.04cm}{\color{red}$\times$}\\
		
		Model represented in \textit{dq} &\hspace{0.07cm}{\color{mygreen}\checkmark}&\hspace{0.07cm}{\color{mygreen}\checkmark}&\hspace{0.07cm}{\color{mygreen}\checkmark}&\hspace{0.07cm}{\color{mygreen}\checkmark}&\hspace{0.04cm}{\color{red}$\times$}&\hspace{0.04cm}{\color{red}$\times$}&\hspace{0.04cm}{\color{red}$\times$}&\hspace{0.07cm}{\color{mygreen}\checkmark}\\ 
		
		Post-fault torque&\hspace{0.07cm}{\color{mygreen}\checkmark}&\hspace{0.07cm}{\color{mygreen}\checkmark}&\hspace{0.04cm}{\color{red}$\times$}&\hspace{0.07cm}{\color{mygreen}\checkmark}&\hspace{0.07cm}{\color{mygreen}\checkmark}&\hspace{0.07cm}{\color{mygreen}\checkmark}&\hspace{0.04cm}{\color{red}$\times$}&\hspace{0.07cm}{\color{mygreen}\checkmark}\\  
		
		Discrete time &\hspace{0.07cm}{\color{mygreen}\checkmark}&\hspace{0.07cm}{\color{mygreen}\checkmark}&\hspace{0.04cm}{\color{red}$\times$}&\hspace{0.04cm}{\color{red}$\times$}&\hspace{0.04cm}{\color{red}$\times$}&\hspace{0.04cm}{\color{red}$\times$}&\hspace{0.04cm}{\color{red}$\times$}&\hspace{0.07cm}{\color{mygreen}\checkmark}\\  
	\end{tabular}\vspace{-0.1cm}
\end{table}
We derived a DTM for an ISC in an interior PMSM with a series-parallel phase segment connection, radial permanent magnet fluxes, and a high connection resistance. In Table~\ref{tab:comparison}, the essential properties of the DTM were compared with other approaches. As shown in Table \ref{tab:comparison}, the model combines several features, the only drawback being the neglected inductive couplings between the phase segments. This limitation reduces the model’s usability for motors where the phase coils are positioned next to each other. Conversely, the DTM exhibits superior precision when the phase segments are interlaced with coils of other phases. Moreover, by employing the proposed discretization method, the derived DTM overcomes the numerical stability problems inherent in the forward Euler approach. Finally, the DTM is computationally efficient, suitable for model-based fault mitigation and diagnostics, and enables rapid generation of simulation datasets with diverse fault indicator values for training diagnostic neural networks.

\section*{Acknowledgment}
L. Zezula is a Brno Ph.D. Talent Scholarship Holder funded by Brno City Municipality.


\vspace{-1.2cm}
\begin{IEEEbiography}[{\includegraphics[width=1in,height=1.25in,clip,keepaspectratio]{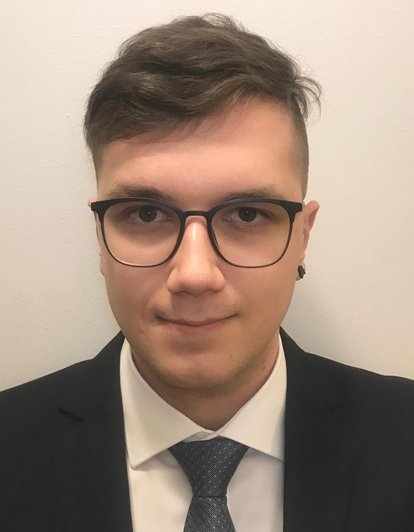}}]
{Lukas Zezula} received the M.Sc. degrees in cybernetics, control and measurements and strategic company development from Brno University of Technology, Brno, the Czech Republic, in 2022 and 2024, respectively. He is currently working towards a Ph.D. in cybernetics, control and measurements.

He is presently a Research Assistant with Central European Institute of Technology, Brno University of Technology. His research interests include the continuous and discrete-time modeling of ac machines under failures, and diagnostics in ac electric motors.
\end{IEEEbiography}

\vspace{-1.2cm}
\begin{IEEEbiography}[{\includegraphics[width=1in,height=1.25in,clip,keepaspectratio]{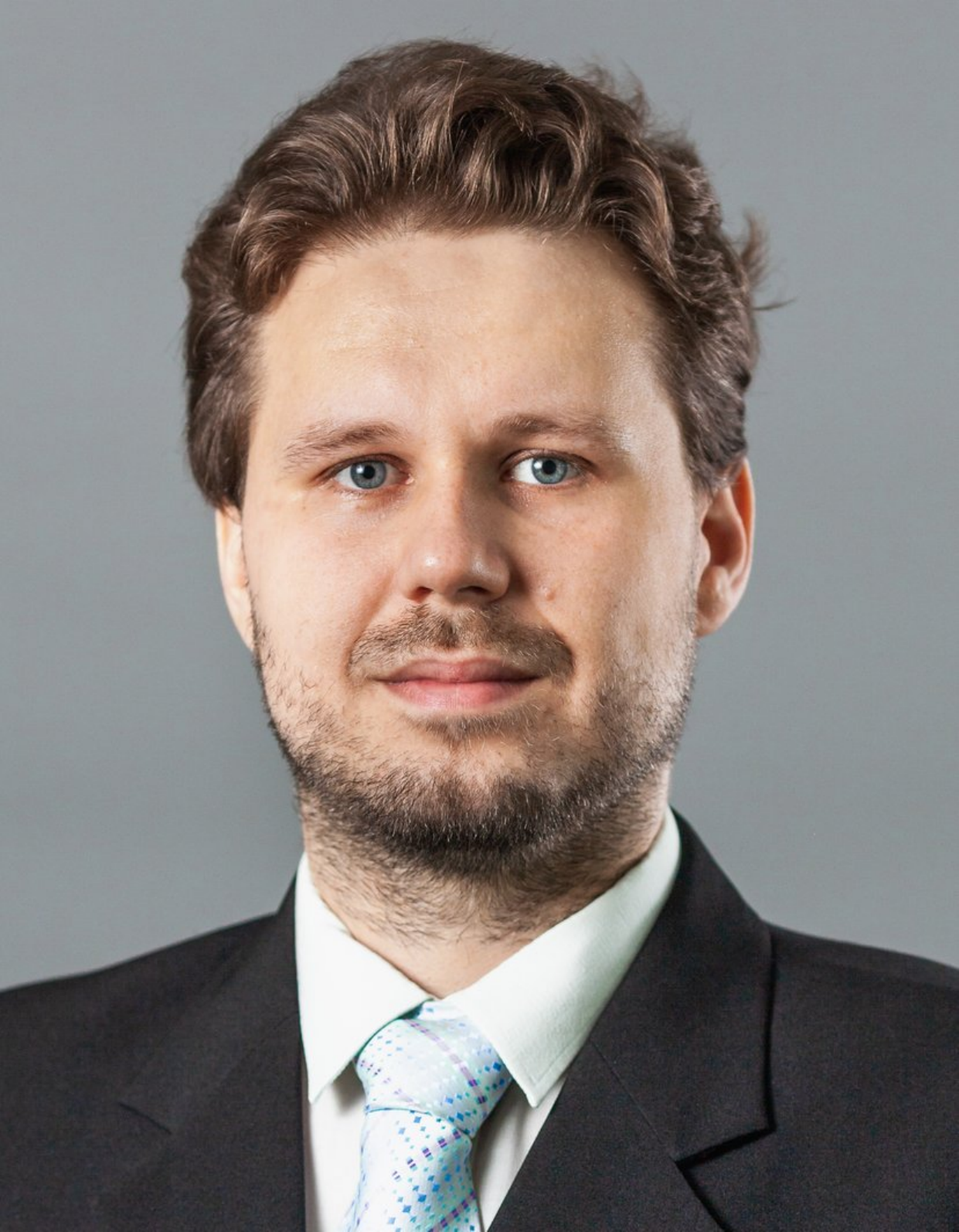}}]
{Matus Kozovsky} received the Ph.D. degree in cybernetics, control and measurements from Brno University of Technology, Brno, the Czech Republic, in 2021.

He is currently a Researcher of SW development with Central European Institute of Technology, Brno University of Technology. His research interests include the fault-tolerant control of ac electric motors, high-speed control of electric motors, and inverter SW improvements.
\end{IEEEbiography}

\vspace{-1.2cm}
\begin{IEEEbiography}[{\includegraphics[width=1in,height=1.25in,clip,keepaspectratio]{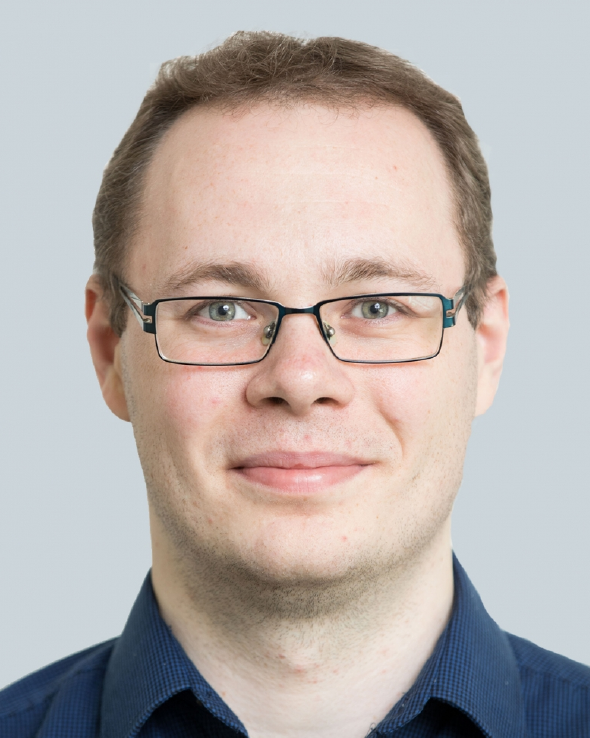}}]
{Ludek Buchta} received the Ph.D. degree in cybernetics, control and measurements from Brno University of Technology, Brno, the Czech Republic, in 2019.
	
He is currently a researcher with Central European Institute of Technology, Brno University of Technology. His research interests include compensating the non-linearity of voltage source inverters, implementing neural network algorithms, and developing advanced algorithms to control AC electric motors.
\end{IEEEbiography}

\vspace{-1.2cm}
\begin{IEEEbiography}[{\includegraphics[width=1in,height=1.25in,clip,keepaspectratio]{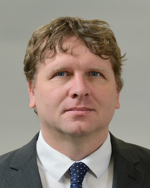}}]
{Petr Blaha} received the M.Sc. degree in cybernetics, control and measurements and a Ph.D. in cybernetics and informatics from Brno University of Technology, Brno, the Czech Republic, in 1996 and 2001, respectively.

He is currently a Senior Researcher with Central European Institute of Technology, Brno University of Technology, Brno, Czech Republic. Since 2007, he has been an Associate Professor at Brno University of Technology. His research interests include the parameter identification and advanced control of ac electric motors and the fault tolerant control of electrical motor drives.
\end{IEEEbiography}

\end{document}